\documentstyle[11pt]{article}
\begin{document}
\begin{titlepage}
\title{\bf Monopole Equations on 8-Manifolds with Spin(7) Holonomy}
\date{ }
\author{ {\bf Ay\c{s}e H\"{u}meyra Bilge}\\{\small
Department of Mathematics}\\
{\small \.{I}stanbul Technical University}\\
{\small \.{I}stanbul, Turkey}\\
{\footnotesize E.mail : bilge@hidiv.cc.itu.edu.tr}\\
{\bf Tekin Dereli}\\{\small Department of Physics}\\
{\small Middle East Technical University}\\{\small Ankara, Turkey}
\\{\footnotesize E.mail : tekin@dereli.physics.metu.edu.tr}\\
{\bf \c{S}ahin Ko\c{c}ak}\\{\small Department of Mathematics}\\
{\small Anadolu University}\\{\small Eski\c{s}ehir, Turkey}\\
{\footnotesize E.mail : skocak@vm.baum.anadolu.edu.tr} }
\maketitle
\begin{abstract}
\noindent {\small We construct a consistent set of monopole equations on 
eight-manifolds with Spin(7) holonomy.
These equations are elliptic and
admit non-trivial solutions including all the 4-dimensional 
Seiberg-Witten solutions as a special case.}
\end{abstract} 
\end{titlepage}

\noindent {\bf 1. Introduction} 
\vskip 4mm

In a remarkable paper ${}^{[1]}$
Seiberg and Witten have shown that diffeomorphism invariants of 4-manifolds 
can be found essentially by counting the
number of solutions of a set of massless, 
Abelian monopole equations ${}^{[2],[3]}$.
It is later noted that 
topological quantum field theories 
which are extensively studied in this context in 2, 3 
and 4 dimensions also exist in 
higher dimensions ${}^{[4],[5],[6],[7]}$. 
Therefore it is of interest to consider 
monopole equations in higher dimensions and thus generalizing the 
4-dimensional Seiberg-Witten theory.

In fact Seiberg-Witten equations 
can be constructed on any even dimensional manifold (D=2n)
with a spin$^c$-structure ${}^{[8]}$. But there are 
problems. The self-duality of 2-forms
plays an eminent role in 4-dimensional theory and we encounter projection 
maps $ \rho^+(F_A)= \rho^+(F_A^+) = \rho (F_A^+)$ 
(see the next section).
The first projection $\rho^+(F_A)$ is 
meaningful in any dimension $2n\geq 4$. However, a straightforward generalization 
of the Seiberg-Witten equations
using this projection
yields an over determined  set of equations having no
non-trivial solutions even locally ${}^{[9]}$.
To use the other projections, one needs an appropriately
generalized notion of self-dual 2-forms.
On the other hand
there is no unique definition of self-duality in higher than four dimensions.
In a previous paper ${}^{[10]}$ we reviewed the existing
definitions of self-duality and gave an eigenvalue
criterion for specifying  self-dual 2-forms on any even dimensional manifold.
In particular, in 
$D=8$ dimensions, there is a linear notion of self-duality defined
on 8-manifolds with Spin(7) holonomy ${}^{[11],[12]}$. 
This corresponds to a specific choice of a maximal linear
subspace in the set of (non-linear) self-dual 2-forms
as defined by our eigenvalue criterion ${}^{[13]}$. 
Eight dimensions is special because in this particular case
the set of linear
Spin(7) self-duality equations can be solved by making use
of octonions ${}^{[14]}$ . 
The existence of {\sl octonionic instantons} 
which realise the last Hopf
fibration $S^{15} \rightarrow S^{8}$ is closely related with the
properties of the octonion algebra ${}^{[15],[16],[17]}$.

Here we use this linear notion of self-duality
to construct a  consistent set of Abelian monopole equations
on 8-manifolds with Spin(7) holonomy. 
These equations turn out to be elliptic and 
locally they admit non-trivial solutions which include  
all 4-dimensional Seiberg-Witten solutions 
as a special case. But before giving our 8-dimensional 
monopole equations,
we first wish in the next section to give
the set up and generalizations of 4-dimensional Seiberg-Witten 
equations to arbitrary
even dimensional manifolds with spin$^c$-structure as proposed by
Salamon ${}^{[8]}$. This is going to help us 
put our monopole equations  
into their proper context. We also wish to note that
any 8-manifold with Spin(7) holonomy is automatically a spin manifold
${}^{[18],[19]}$ 
and thus carries a spin$^c$-structure; making the application 
of the general approach  possible.
In fact our monopole equations can always be expressed purely in the 
real realm, but in order to relate them to the 
4-dimensional Seiberg-Witten equations,
it is preferable to use the spin$^{c}$-structure and complex spinors.
\vskip 4mm
\noindent {\bf 2. Definitions and notation}
\vskip 4mm

A spin$^c$-structure on a $2n$-dimensional real inner-product space $V$ is
a pair $(W,\Gamma)$, where $W$ is a $2^n$-dimensional complex Hermitian space
and  $\Gamma  : V \rightarrow End(W)$ is a linear map satisfying
$${\Gamma(v)}^*=-\Gamma(v),\qquad {\Gamma(v)}^2=-{\Vert v \Vert}^2$$
for $v\in V$. Globalizing this defines the notion of a spin$^c$-structure
$\Gamma :
 TX \rightarrow End(W)$ on a $2n$-dimensional 
(oriented) manifold $X$,
$W$ being a $2^n$-dimensional complex Hermitian vector bundle on $X$. 
Such a structure exists if and only if
$w_2(X)$ has an integral lift. $\Gamma$ extends 
to an isomorphism between the complex Clifford algebra bundle $C^c(TX)$ and 
$End(W)$.
 There is a natural splitting $W=W^+ \oplus W^-$
into the ${\pm}i^n$ eigenspaces of $\Gamma(e_{2n}e_{2n-1}\cdots{e_1})$ 
where $e_1,e_2,\cdots,{e_{2n}}$ is any positively oriented local orthonormal 
frame of $TX$.

The extension of $\Gamma$ to $C_{2}(X)$ gives, via the identification of
$\Lambda^{2}(T^{*}X)$ with $C_{2}(X)$, a map
$$\rho  : \Lambda^{2}(T^{*}X) \rightarrow End(W)$$
given by
$$\rho(\sum_{i<j}\eta_{ij}e_i^*\wedge e_j^*)=
\sum_{i<j}\eta_{ij}\Gamma(e_i)\Gamma(e_j).$$
The bundles $W^\pm$ are  invariant under $\rho(\eta)$ for 
$\eta\in\Lambda^{2}(T^{*}X)$.
Denote $\rho^\pm (\eta)=\rho(\eta)\vert_{W^\pm}$. 
The map $\rho$ (and $\rho^\pm$)
extends to
$$\rho:\Lambda^{2}(T^{*}X)\otimes{\bf C}\rightarrow End(W).$$
(If $\eta\in\Lambda^{2}(T^{*}X)\otimes\bf C$ 
is real-valued then $\rho(\eta)$
is skew-Hermitian and if $\eta$ is imaginary-valued then 
$\rho(\eta)$ is Hermitian.)
A Hermitian connection $\nabla$ on $W$ is called a spin$^c$ connection 
(compatible with the Levi-Civita connection) if
$$\nabla_v(\Gamma(w)\Phi)=\Gamma(w)\nabla_v\Phi+\Gamma(\nabla_vw)\Phi$$
where $\Phi$ is a spinor (section of $W$), $v$ and $w$ are vector fields 
on $X$ and $\nabla_vw$ is the Levi-Civita connection on $X$. 
$\nabla$ preserves the subbundles $W^\pm$.
There is a principal Spin$^c(2n)=\lbrace e^{i\theta}x\vert \theta\in {\bf R}, x\in Spin(2n) \rbrace
\subset C^c ({\bf R}^{2n})$ bundle  $P$ on $X$ such that $W$ and $TX$ can be
recovered as the associated bundles
$$W=P\times_{Spin^c(2n)}{\bf C}^{2^n}, \qquad TX=P\times_{Ad}{\bf R}^{2n},$$
$Ad$ being the adjoint action of Spin$^c(2n)$ on ${\bf R}^{2n}$. We get then
a complex line bundle $L_{\Gamma}=P\times_{\delta}{\bf C}$ using the map
$\delta:Spin^c(2n)\rightarrow S^1$ given by $\delta ( e^{i\theta}x )
=e^{2i\theta}$.

There is a one-to-one correspondence between spin$^c$ connections on $W$ and
spin$^c$(2n)=Lie(Spin$^c$(2n)=spin(2n)$\oplus i{\bf R}$ -valued 
connection 1-forms $ \hat A\in { \bf A} (P)
\subset \Omega^1 (P$,spin$^c$(2n)) on $P$.
Now consider the trace-part $A$ of $\hat A$: $A=\frac{1}{2^n}trace(\hat A)$.
This is an imaginary valued 1-form $A\in \Omega^1 (P,i\bf R)$ 
which is equivariant
and satisfies $$A_p(p\cdot \xi)=\frac{1}{2^n}trace(\xi)$$ for 
$v\in T_pP,g\in$ 
Spin$^c$(2n),$ \xi \in$ spin$^c$(2n) (where $p\cdot \xi$ is the 
infinitesimal action).
Denote the set of imaginary valued 1-forms on $P$ satisfying these two 
properties by ${\bf A}(\Gamma)$. There is a one-to-one correspondence between
these 1-forms and spin$^c$ connections on $W$. Denote the connection
corresponding to $A$ by $\nabla_A$. ${\bf A}(\Gamma)$ is an affine
space with parallel vector space $\Omega^1 (X,i\bf R)$. 
For $A\in {\bf A}(\Gamma)$, the 
1-form $2A\in \Omega^1 (P,i\bf R)$ represents a connection on the line
bundle $L_{\Gamma}$. Because of this reason $A$ is called a {\it virtual
connection} on the {\it virtual line bundle} ${L_{\Gamma}^{1/2}}$.
Let $F_A \in\Omega^2 (X,i\bf R)$ denote the curvature of the 1-form $A$.
Finally, let $D_A$ denote the Dirac operator corresponding to $A\in {\bf A}
(\Gamma)$,
$$D_A : C^\infty (X,W^+)\rightarrow C^\infty (X,W^-)$$
defined by
$$D_A (\Phi)=\sum_{i=1}^{2n}{\Gamma(e_i){\nabla_{A,e_i}}}(\Phi)$$
where $\Phi \in C^\infty (X,W^+)$ and $e_1,e_2,\cdots,{e_{2n}}$ is any local
orthonormal frame.

The Seiberg-Witten equations can now be expressed as follows. 
Fix a spin$^c$-structure $\Gamma:TX\rightarrow End(W)$ on $X$ and consider the pair
$(A,\Phi)\in {{\bf A}(\Gamma) \times C^\infty (X,W^+)}$. The Seiberg-
Witten equations read
$$D_A(\Phi)=0\ \ , \qquad \rho^+(F_A)=(\Phi \Phi^*)_0$$
where $(\Phi \Phi^*)_0 \in C^\infty (X,End(W^+))$
is defined by $(\Phi \Phi^*)(\tau)=<\Phi, \tau> \Phi$ for $\tau \in C^\infty 
(X,W^+)$ and $(\Phi \Phi^*)_0$ is the traceless part of $(\Phi \Phi^*)$.
\vskip 0.4cm
\noindent {\bf 3. Seiberg-Witten equations on 4-manifolds}
\vskip .4cm
Before going over to 8-manifolds, we first show that the
Seiberg-Witten equations on 4-manifolds (Ref.[8], p.232) can be rewritten 
in a different form. The Dirac equation
\begin{equation}
D_A(\Phi)=0
\end{equation}
can be explicitly written as
\begin{equation}
\nabla_1 \Phi =  I \nabla_2 \Phi + J \nabla_3 \Phi +K \nabla_4 \Phi,
\end{equation}
and 
\begin{equation}
\rho^+(F_A)=(\Phi \Phi^*)_0
\end{equation}
 is equivalent to the set
\begin{eqnarray}
F_{12}+F_{34}&=&-1/2 {\Phi}^* I \Phi, \nonumber \\
F_{13}-F_{24}&=&-1/2 {\Phi}^* J \Phi, \nonumber \\
F_{14}+F_{23}&=&-1/2 {\Phi}^* K \Phi,
\end{eqnarray}
where 
$\Phi : {\bf R}^4 \rightarrow {\bf C}^2$, $\nabla_i{\Phi}= \frac {\partial \Phi}{\partial x_i}+ A_i \Phi$,\\
$A=\sum_{i=1}^4 A_i dx_i \in {\Omega}^1 ({\bf R}^4,i{\bf R})$,
$F_A=\sum_{i<j}F_{ij} dx_i \wedge dx_j \in {\Omega}^2 ({\bf R}^4,i{\bf R})$,
 
\noindent and

$I= \left [ \begin{array}{cc} i & 0 \cr
              0 & -i  \end{array} \right ]$,
$J=\left [ \begin{array}{cc} 0 & 1 \cr
              -1 & 0  \end{array} \right ]$,
$K=\left [ \begin{array}{cc} 0 & i \cr
              i & 0  \end{array} \right ]$.\\
\vskip 3mm

\noindent In the most explicit form, these equations can be written as
\begin{eqnarray}
\frac {\partial \phi_1}{\partial x_1}+A_1 \phi_1&=&
i(\frac {\partial \phi_1}{\partial x_2}+A_2 \phi_1)+
\frac {\partial \phi_2}{\partial x_3}+A_3 \phi_2+
i(\frac {\partial \phi_2}{\partial x_4}+A_4 \phi_2), \nonumber \\
\frac {\partial \phi_2}{\partial x_1}+A_1 \phi_2 &=&
-i(\frac {\partial \phi_2}{\partial x_2}+A_2 \phi_2)
-(\frac {\partial \phi_1}{\partial x_3}+A_3 \phi_1)+
i(\frac {\partial \phi_1}{\partial x_4}+A_4 \phi_1)
\end{eqnarray}
(for $D_A(\Phi)=0$) and
\begin{eqnarray}
F_{12}+F_{34}&=&-i/2(\phi_1 {\bar \phi_1}-\phi_2 {\bar \phi_2}), \nonumber \\
F_{13}-F_{24}&=&1/2(\phi_1 {\bar \phi_2}-\phi_2 {\bar \phi_1}), \nonumber \\
F_{14}+F_{23}&=&-i/2(\phi_1 {\bar \phi_2}+\phi_2 {\bar \phi_1})
\end{eqnarray}
(for $\rho^+(F_A)=(\Phi \Phi^*)_0$).

We will reinterpret the second part of these equations in the following way:
The 6-dimensional bundle of real-valued 2-forms on ${\bf R}^4$ has a 
3-dimensional subbundle of self-dual forms with orthogonal basis
\begin{eqnarray}
f_1&=&dx_1\wedge dx_2+dx_3\wedge dx_4, \nonumber \\
f_2&=&dx_1\wedge dx_3-dx_2\wedge dx_4, \nonumber \\
f_3&=&dx_1\wedge dx_4+dx_2\wedge dx_3, 
\end{eqnarray}
in each fiber with respect to the usual metric.
These forms span a 3-dimensional complex subbundle of the bundle of 
complex-valued 2-forms. The projection of a (global) 2-form 
$F=\sum F_{ij} dx_i \wedge dx_j\in {\Omega}^2 ({\bf R}^4,i{\bf R})$
onto this complex subbundle is given by  
\begin{equation}
F^+= 1/2(F_{12}+F_{34})f_1+
1/2(F_{13}-F_{24})f_2+1/2(F_{14}+F_{23})f_3 .
\end{equation}
We have $\rho^+ (f_1)= 2I, \rho^+ (f_2)= 2J, \rho^+ (f_3)= 2K$, so that,
\begin{equation}
\rho^+(F^+)=(F_{12}+F_{34})I+(F_{13}-F_{24})J+(F_{14}+F_{23})K .
\end{equation}
On the other hand, the orthogonal projection $(\Phi \Phi^*)^+$ of 
$\Phi \Phi^*$ onto the subbundle of the positive spinor bundle generated by 
the (Hermitian-) orthogonal basis
$(\rho^+ (f_1), \rho^+ (f_2), \rho^+ (f_3))$
is given by
$$
<2I, \Phi \Phi^*>2I/{\vert 2I \vert}^2
+<2J, \Phi \Phi^*>2J/{\vert 2J \vert}^2
+<2K, \Phi \Phi^*>2K/{\vert 2K \vert}^2$$
\begin{equation} 
= \frac{1}{2}<I, \Phi \Phi^*>I+ \frac{1}{2}<J, \Phi \Phi^*>J+
\frac{1}{2}<K, \Phi \Phi^*>K. 
\end{equation}

\noindent Since 
\begin{equation}
 <I, \Phi \Phi^*>=- \Phi^* I \Phi,\quad 
<J, \Phi \Phi^*>=- \Phi^* J \Phi,\quad
<K, \Phi \Phi^*>=- \Phi^* K \Phi,
\end{equation}
this shows that the second part of the Seiberg-Witten equations 
can be expressed as
follows: Given any (global, imaginary-valued) 2-form $F$, the image under 
the map $\rho^+$ of its self-dual part $F^+$ coincides with the orthogonal  
projection of $\Phi \Phi^*$ onto the subbundle of the positive spinor bundle 
which is the image bundle of the complexified subbundle of self-dual 2-forms 
under the map $\rho^+$,  that is, 
\begin{equation}
\rho^+(F^+)=(\Phi \Phi^*)^+.
\end{equation}
Indeed, in the present case $(\Phi \Phi^*)^+$ is nothing else than 
$(\Phi \Phi^*)_0.$
In this modified form the Seiberg-Witten equations allow a tempting 
generalisation.
Suppose we are given a subbundle $S \subset \Lambda^{2}(T^{*}X)$.
Denote the complexification  of $S$ by $S^*$, the projection of an 
imaginary valued 2-form field $F$ onto $S^*$ by $F^+$ and the projection of
$\phi \phi^*$ onto $\rho^{+}(S^*)$ by $(\phi \phi^*)^+$. 
Then the equation $\rho^{+}(F^+)= (\phi \phi^*)^+$ can be taken as a
substitute of the 4-dimensional equation (3) in 2n-dimensions.
An arbitrary choice of $S$ wouldn't probably give anything interesting,
but stable subbundles with respect to certain structures on $X$ are
likely to give useful equations.

\vskip 0.4cm
\noindent {\bf 4. Monopole equations on 8-manifolds}
\vskip .4cm

We now consider 8-manifolds with Spin(7) holonomy. In this case
there are two natural choices of $S$
which have already found applications in the  existing
literature.
In the 28-dimensional space of 2-forms $\Omega^{2}({\bf R}^{8},{\bf R})$,
there are two orthogonal subspaces $S_{1}$ and $S_2$ 
( 7 and 21 dimensional, respectively)  which are 
Spin(7) $\subset$ SO(8) invariant ${}^{[11],[12]}$.
On an 8-manifold $X$ with Spin(7) holonomy
(so that the structure group is reducible to Spin(7)) they give rise to global 
subbundles (denoted by the same letters) 
$S_1 , S_2 \subset \Lambda^{2}(T^{*}X)$
which can play the above mentioned role. We will concentrate
on the 7-dimensional subbundle $S_1$ and show that the resulting equations 
are elliptic, exemplify the local existence of non-trivial solutions
and show that they are related to solutions of the 
4-dimensional Seiberg-Witten equations. We would like to point out that
instead of the widely known CDFN 7-plane,  we are  working with
another 7-plane in $\Omega^{2}({\bf R}^{8},{\bf R})$, which is
conjugated to the CDFN 7-plane and thus invariant under
a conjugated Spin(7) embedding in SO(8).
This has the advantage that the 2-forms in this 7-plane can be expressed
in an elegant way in terms of
4-dimensional self-dual and anti-self-dual 2-forms.
(For a general account we refer to our previous work, Ref.[10].)
We will define this 7-plane below, but before that, for the sake of clarity,
we first wish to present the global monopole equations.   
Let $X$ be an 8-manifold with Spin(7) holonomy
 and $S$ be any stable subbundle of $\Lambda^{2}(T^{*}X)$ and $S^*$ its
complexification.
Given an imaginary valued global 2-form $F$, let us denote its projection 
onto
$S^*$ by $F^+$  and the projection of any global spinor $\phi$ onto the
subbundle $\rho^{+}(S^*) \subset End(W^+)$ by $\phi^{+}$.
Then the monopole equations read
\begin{equation}
D_{A}(\phi)=0,
\end{equation}
\begin{equation} 
\rho^{+}(F_{A}^{+})=(\phi \phi^{*})^{+}.
\end{equation}
\vskip 5mm

Now, we define $S_1 \subset \Omega^{2}({\bf R}^{8},{\bf R})$ to be the 
linear space of 2-forms
$$\omega=\sum_{i<j}\omega_{ij} dx_i \wedge dx_j \in {\Omega}^2 
({\bf R}^8,{\bf R}),$$
which can be expressed in matrix form as
\begin{equation}
\omega = \omega_{12} f +\pmatrix{\omega'&\omega''\cr
                     \omega''&-\omega'\cr},
\end{equation}
where $\omega_{12}$ is a real function,
$\omega'$ is the matrix of a 4-dimensional self-dual 2-form,
$\omega''$ is the matrix of a 4-dimensional anti-self-dual
2-form and we let $f= -J \otimes id_{4}$.
These 2-forms span a 7-dimensional linear
subspace $S_1$ in the 28-dimensional space of 2-forms and the square
of any element in this subspace is a scalar matrix. $S_1$ is 
maximal with respect to this property.
We choose the following orhogonal basis for this 
maximal linear subspace of self-dual 2-forms:
$$f_1=dx_1 \wedge dx_5 + dx_2 \wedge dx_6 +dx_3 
\wedge dx_7 +dx_4 \wedge dx_8,$$
$$f_2=dx_1 \wedge dx_2 + dx_3 \wedge dx_4 -dx_5 \wedge 
dx_6 -dx_7 \wedge dx_8,$$
$$f_3=dx_1 \wedge dx_6 - dx_2 \wedge dx_5 -dx_3 \wedge 
dx_8 +dx_4 \wedge dx_7,$$
$$f_4=dx_1 \wedge dx_3 - dx_2 \wedge dx_4 -dx_5 \wedge dx_7 +
dx_6 \wedge dx_8,$$
$$f_5=dx_1 \wedge dx_7 + dx_2 \wedge dx_8 -dx_3 \wedge 
dx_5 -dx_4 \wedge dx_6,$$
$$f_6=dx_1 \wedge dx_4 + dx_2 \wedge dx_3 -dx_5 \wedge dx_8 -
dx_6 \wedge dx_7,$$
\begin{equation}
f_7=dx_1 \wedge dx_8 - dx_2 \wedge dx_7 +dx_3 \wedge 
dx_6 -dx_4 \wedge dx_5.
\end{equation}
In  matrix notation we set  $f_1=f $, and take 
$$f_2= -iI \otimes a_{1} , \hskip 6mm
f_3= -iK \otimes b_{1}$$ 
$$f_4= -iI \otimes a_{2} , \hskip 6mm
f_5= -iK \otimes b_{2}$$
\begin{equation}
f_6= -iI \otimes a_{3} , \hskip 6mm 
f_7= -iK \otimes b_{3} .
\end{equation}

\noindent where $( I, J, K )$ are as given as before and we have 
$$ a_{1} =\left ( \begin{array}{cccc}
0&-1&0&0\\1&0&0&0\\0&0&0&-1\\0&0&1&0 \end{array} \right ) ,
a_{2} = \left ( \begin{array}{cccc}
0&0&-1&0\\0&0&0&1\\1&0&0&0\\0&-1&0&0 \end{array} \right ) ,
a_{3} = \left ( \begin{array}{cccc} 0&0&0&1 \\0&0&1&0\\0&-1&0&0\\-1&0&0&0
\end{array} \right )$$
\noindent and 
$$ b_{1}= \left ( \begin{array}{cccc} 0&-1&0&0\\1&0&0&0\\0&0&0&1\\0&0&-1&0
\end{array} \right )  ,
b_{2}= \left ( \begin{array}{cccc} 0&0&-1&0\\
0&0&0&-1\\1&0&0&0\\ 0&1&0&0 \end{array} \right )  ,
b_{3}= \left ( \begin{array}{cccc} 0&0&0&1\\0&0&-1&0\\
0&1&0&0\\ -1&0&0&0 \end{array} \right ) .$$

At this point it will be instructive to show that the above basis
corresponds to a representation of the Clifford algebra $\sl Cl_7$
induced by right multiplications in the algebra of octonions. We adopt 
the Cayley-Dickson approach and describe
a quaternion by a pair of complex numbers so that
$ a = (x+iy)+j(u+iv)$ where $(i,j,ij=k)$ are the imaginary unit
quaternions. In a similar way an octonion is described
by a pair of quaternions $(a,b)$. Then the octonionic multiplication rule is
\begin{equation}
 (a,b) \cdot (c,d) = (ac-\bar{d}b, da+b\bar{c}).
\end{equation}
If we now represent an octonion $(a,b)$ by a vector in ${\bf R}^8$,
its right multiplication by imaginary unit octonions correspond to
linear transformations on ${\bf R}^8$. We thus obtain the
following correspondences:
$$ (0,1) \rightarrow f_1 , \\ (i,0) \rightarrow f_2  ,  (j,0) \rightarrow f_3 , 
(k,0) \rightarrow f_4, $$
\begin{equation}
(0,i) \rightarrow f_5  , (0,j) \rightarrow f_6  , (0,k) \rightarrow f_7 .\\
\end{equation}    

\vskip 3mm 

The projection $F^+$ of a 2-form 
$F_=\sum_{i<j}F_{ij} dx_i \wedge dx_j\in {\Omega}^2 ({\bf R}^8,i{\bf R})$
onto the complexification of the above self-dual subspace is given by

$$F^+=1/4(F_{15}+F_{26}+F_{37}+F_{48})f_1$$
$$ \qquad \ \   + 1/4(F_{12}+F_{34}-F_{56}-F_{78})f_2$$
$$ \qquad \ \   + 1/4(F_{16}-F_{25}-F_{38}+F_{47})f_3$$
$$ \qquad \ \   + 1/4(F_{13}-F_{24}-F_{57}+F_{68})f_4$$
$$ \qquad \ \   + 1/4(F_{17}+F_{28}-F_{35}-F_{46})f_5$$
$$ \qquad \ \   + 1/4(F_{14}+F_{23}-F_{58}-F_{67})f_6$$
$$ \qquad \ \   + 1/4(F_{18}-F_{27}+F_{36}-F_{45})f_7.$$

\noindent We now fix the constant spin$^c$-structure $\Gamma:{\bf R}^{\bf 8} 
\longrightarrow {\bf C}^{{\bf 16}\times{\bf 16}}$ given by 
\begin{equation}
\Gamma(e_i)=\pmatrix{0&\gamma(e_i)\cr -{\gamma(e_i)}^{*}&0}
\end{equation}
where ${e_i} ,i=1,2,...,8$ is the standard basis for ${\bf R}^{\bf 8}$ and
$\gamma(e_1)=Id, \quad \gamma(e_i)=f_{i-1}$  for $i=2,3,...,8.$
We note that this choice is specific to 8 dimensions , because 
$2n=2^{n-1}$ only for $n=4.$  
We have $X={\bf R}^8, 
W={\bf R}^8 \times {\bf C}^{16}, 
W^\pm={\bf R}^8 \times {\bf C}^8$ and 
$L_\Gamma = {L_\Gamma}^{1/2}={\bf R}^8 \times {\bf C}$.
Consider the connection 1-form 
\begin{equation}
A=\sum_{i=1}^8 A_i dx_i \in {\Omega}^1 ({\bf R}^8,i{\bf R})
\end{equation}
on the line bundle ${\bf R}^8 \times {\bf C}$. Its curvature is given by
\begin{equation}
F_A=\sum_{i<j}F_{ij} dx_i \wedge dx_j \in {\Omega}^2 ({\bf R}^8,i{\bf R})
\end{equation}
where 
$F_{ij}=\frac {\partial A_j}{\partial x_i} - \frac {\partial A_i}{\partial 
x_j}$. The spin$^c$ connection $\nabla= \nabla_A$ on $W^+$ is given by
\begin{equation}
\nabla_i{\Phi}= \frac {\partial \Phi}{\partial x_i}+ A_i \Phi
\end{equation}
($i=1,...,8$) where $\Phi : {\bf R}^8 \rightarrow {\bf C}^8$. Therefore the map 
$$\rho^+:\Lambda^{2}(T^{*}X)\otimes{\bf C}\rightarrow End(W^+)$$
can be computed for our generators $f_i$ to give

$$\rho^+(f_1)=\gamma (e_1) \gamma (e_5)+\gamma (e_2) \gamma (e_6)+
\gamma (e_3) \gamma (e_7)+\gamma (e_4) \gamma (e_8)$$
$$\rho^+(f_2)=\gamma (e_1) \gamma (e_2)+\gamma (e_3) \gamma (e_4)-
\gamma (e_5) \gamma (e_6)-\gamma (e_7) \gamma (e_8)$$
$$\rho^+(f_3)=\gamma (e_1) \gamma (e_6)-\gamma (e_2) \gamma (e_5)+
\gamma (e_3) \gamma (e_8)+\gamma (e_4) \gamma (e_7)$$
$$\rho^+(f_4)=\gamma (e_1) \gamma (e_3)-\gamma (e_2) \gamma (e_4)-
\gamma (e_5) \gamma (e_7)+\gamma (e_6) \gamma (e_8)$$
$$\rho^+(f_5)=\gamma (e_1) \gamma (e_7)+\gamma (e_2) \gamma (e_8)-
\gamma (e_3) \gamma (e_5)-\gamma (e_4) \gamma (e_6)$$
$$\rho^+(f_6)=\gamma (e_1) \gamma (e_4)+\gamma (e_2) \gamma (e_3)-
\gamma (e_5) \gamma (e_8)-\gamma (e_6) \gamma (e_7)$$
$$\rho^+(f_7)=\gamma (e_1) \gamma (e_8)-\gamma (e_2) \gamma (e_7)+
\gamma (e_3) \gamma (e_6)-\gamma (e_4) \gamma (e_5).$$
\vskip 2mm
\noindent Then for a connection $A=\sum_{i=1}^8 A_i dx_i \in {\Omega}^1 ({\bf R}^8,i{\bf R})$
and a given  complex 8-spinor
$\Psi=(\psi_1,\psi_2,...,\psi_8) \in C^\infty(X,W^+)=C^\infty ({\bf R}^8,
{\bf R}^8\times {\bf C}^8)$
we  state our 8-dimensional monopole equations as follows:
\begin{equation}
D_A(\Psi)=0\ \ , \qquad \rho^+({F_A}^+)=(\Psi \Psi^*)^+ .
\end{equation}

\noindent Here $(\Psi \Psi^*)^+$ is the orthogonal projection of $\Psi \Psi^*$ 
onto the spinor subbundle spanned by $\rho^+(f_i), i=1,2,...,7$.
More explicitly, $D_A(\Psi)=0$ can be expressed as
\begin{equation}
\nabla_1 \Psi =  \gamma(e_2) \nabla_2 \Psi + \gamma(e_3)  \nabla_3 \Psi +
...+\gamma(e_8) \nabla_8 \Psi
\end{equation}
and  $\rho^+({F_A}^+)=(\Psi \Psi^*)^+$ is equivalent to the equation
\begin{equation}
\rho^+({F_A}^+)
=\sum_{i=2} ^8 <\rho^+(f_i),\Psi \Psi^*> \rho^+(f_i)/{\vert \rho^+(f_i) \vert}^2.
\end{equation}
(26) is equivalent to the set of equations

$$ F_{15}+F_{26}+F_{37}+F_{48}=1/8<\rho^+(f_1),\Psi \Psi^*> ,$$
$$ F_{12}+F_{34}-F_{56}-F_{78}=1/8<\rho^+(f_2),\Psi \Psi^*>,$$
$$ F_{16}-F_{25}-F_{38}+F_{47}=1/8<\rho^+(f_3),\Psi \Psi^*>,$$
$$ F_{13}-F_{24}-F_{57}+F_{68}=1/8<\rho^+(f_4),\Psi \Psi^*>,$$
$$ F_{17}+F_{28}-F_{35}-F_{46}=1/8<\rho^+(f_5),\Psi \Psi^*>,$$
$$ F_{14}+F_{23}-F_{58}-F_{67}=1/8<\rho^+(f_6),\Psi \Psi^*>,$$
$$ F_{18}-F_{27}+F_{36}-F_{45}=1/8<\rho^+(f_7),\Psi \Psi^*>.$$
or still more explicitly to the equations
$$
F_{15}+F_{26}+F_{37}+F_{48}=1/4(\psi_1 {\bar \psi }_3- \psi_3 {\bar \psi }_1
-\psi_2 {\bar \psi }_4+\psi_4 {\bar \psi }_2
-\psi_5 {\bar \psi }_7+\psi_7 {\bar \psi }_5
-\psi_6 {\bar \psi }_8+\psi_8 {\bar \psi }_6),$$
$$
F_{12}+F_{34}-F_{56}-F_{78}=1/4(\psi_1 {\bar \psi }_5- \psi_5 {\bar \psi }_1
-\psi_2 {\bar \psi }_6+\psi_6 {\bar \psi }_2
+\psi_3 {\bar \psi }_7-\psi_7 {\bar \psi }_3
+\psi_4 {\bar \psi }_8-\psi_8 {\bar \psi }_4),$$
$$
F_{16}-F_{25}-F_{38}+F_{47}=1/4(\psi_1 {\bar \psi }_7- \psi_7 {\bar \psi }_1
+\psi_2 {\bar \psi }_8-\psi_8 {\bar \psi }_2
-\psi_3 {\bar \psi }_5+\psi_5 {\bar \psi }_3
+\psi_4 {\bar \psi }_6-\psi_6 {\bar \psi }_4),$$
$$
F_{13}-F_{24}-F_{57}+F_{68}=1/4(\psi_1 {\bar \psi }_2- \psi_2 {\bar \psi }_1
+\psi_3 {\bar \psi }_4-\psi_4 {\bar \psi }_3
+\psi_5 {\bar \psi }_6-\psi_6 {\bar \psi }_5
-\psi_7 {\bar \psi }_8+\psi_8 {\bar \psi }_7),$$
$$
F_{17}+F_{28}-F_{35}-F_{46}=1/4(\psi_1 {\bar \psi }_4- \psi_4 {\bar \psi }_1
+\psi_2 {\bar \psi }_3-\psi_3 {\bar \psi }_2
-\psi_5 {\bar \psi }_8+\psi_8 {\bar \psi }_5
+\psi_6 {\bar \psi }_7-\psi_7 {\bar \psi }_6),$$
$$
F_{14}+F_{23}-F_{58}-F_{67}=1/4(-\psi_1 {\bar \psi }_6+ \psi_6 {\bar \psi }_1
-\psi_2 {\bar \psi }_5+\psi_5 {\bar \psi }_2
-\psi_3 {\bar \psi }_8+\psi_8 {\bar \psi }_3
+\psi_4 {\bar \psi }_7-\psi_7 {\bar \psi }_4),$$
$$
F_{18}-F_{27}+F_{36}-F_{45}=1/4(\psi_1 {\bar \psi }_8- \psi_8 {\bar \psi }_1
-\psi_2 {\bar \psi }_7+\psi_7 {\bar \psi }_2
-\psi_3 {\bar \psi }_6+\psi_6 {\bar \psi }_3
-\psi_4 {\bar \psi }_5+\psi_5 {\bar \psi }_4).$$

\newpage
\vskip 0.4cm
\noindent {\bf 5. Conclusion}

We will now show that 
the  system of monopole equations (25)-(26) form an elliptic system.
These equations 
can be written compactly in the form
$$
\langle F,f_i\rangle=1/8\langle \rho^+(f_i),\Psi \Psi^*\rangle,
\quad i=1\dots 7,\quad
D_A(\Psi)=0.
$$
If in addition we impose the Coulomb gauge condition
$$
\sum_{i=1}^8 \partial_i A_i=0,
$$
we obtain  a system of first order partial differential
equations consisting of eight  equations for the
 components of the spinor  $\Psi$ and eight equations for
the  components of the
connection 1-form $A$. The characteristic determinant of this system 
${}^{[20]}$ is  the product of the characteristic determinants
 of the equations for $\Psi$ and $A$.
As the  Dirac operator is elliptic${}^{[19]}$, the ellipticity of the
present system depends on the characteristic determinant of the system
consisting of
$\langle F,f_i\rangle=1/8\langle \rho^+(f_i),\Psi \Psi^*\rangle,
\quad i=1\dots 7$ and the Coulomb gauge condition.
In the computation of the characteristic determinant, the fifth row, 
for instance, is obtained from
$$ F_{15}+F_{26}+F_{37}+F_{48}=\partial_1 A_5 - \partial_5 A_1
+ \partial_2 A_6 - \partial_6 A_2 + \partial_3 A_7 - \partial_7 A_3
+\partial_4 A_8 - \partial_8 A_4$$
by replacing $ \partial_i$ by $ \xi_i$. Thus after a 
rearrangement of the order of the equations, the characteristic determinant
can be obtained as
$$
 det \left (
\begin{array}{cccccccc}
 \xi_1& \xi_2& \xi_3& \xi_4& \xi_5& \xi_6& \xi_7& \xi_8\\
-\xi_2& \xi_1&-\xi_4& \xi_3& \xi_6&-\xi_5& \xi_8&-\xi_7\\
-\xi_3& \xi_4& \xi_1&-\xi_2& \xi_7&-\xi_8&-\xi_5& \xi_6\\
-\xi_4&-\xi_3& \xi_2& \xi_1& \xi_8& \xi_7&-\xi_6&-\xi_5\\
-\xi_5&-\xi_6&-\xi_7&-\xi_8& \xi_1& \xi_2& \xi_3& \xi_4\\
-\xi_6& \xi_5& \xi_8&-\xi_7&-\xi_2& \xi_1& \xi_4&-\xi_3\\
-\xi_7&-\xi_8& \xi_5& \xi_6&-\xi_3&-\xi_4& \xi_1& \xi_2\\
-\xi_8& \xi_7&-\xi_6& \xi_5&-\xi_4& \xi_3&-\xi_2& \xi_1
\end{array} \right ). 
$$
It is equal to
$$(\xi_1^2+\xi_2^2+\xi_3^2+\xi_4^2+\xi_5^2+\xi_6^2+\xi_7^2+\xi_8^2)^4.$$
and this proves ellipticity.
\newpage

Finally we point out that the monopole equations 
(25)-(26) admit non-trivial solutions.
For example, if the pair $(A,\Phi)$ with 
$$A=\sum_{i=1}^4 A_i(x_1,x_2,x_3,x_4) dx_i$$ and 
$$\Phi=(\phi_1(x_1,x_2,x_3,x_4), \phi_2(x_1,x_2,x_3,x_4))$$ 
is a solution of the 4-dimensional Seiberg-Witten equations, 
then the pair $(B,\Psi)$ with 
$$B=\sum_{i=1}^4 A_i(x_1,x_2,x_3,x_4) dx_i$$
(i.e. the first four components $B_i$ of $B$ coincide with $A_i$, thus not
depending on $x_5,x_6,x_7,x_8$ and the last four components of $B$ vanish)
and
$$\Psi=(0,0,\phi_1,\phi_2,0,0,
i\phi_1,-i\phi_2),$$
where $\phi_1$ and $\phi_2$ depend only on 
$x_1, x_2, x_3, x_4$,
is a solution of these new 8-dimensional monopole equations. 
It can directly be verified
that $\Psi$ is harmonic with respect to $B$ and the second part of the 
equations is also satisfied.

\end{document}